\documentclass{article}
\usepackage[utf8]{inputenc}
\usepackage{amsfonts, amssymb, amsthm, environ, etex, etoolbox, lscape, mathtools, MnSymbol, stmaryrd, subcaption, thmtools, tikz, url}

\usepackage[citestyle = authoryear, natbib = true, maxcitenames=2, maxbibnames=99]{biblatex}
\usepackage[bookmarks, bookmarksnumbered, linktocpage, urlcolor = cyan]{hyperref}
\usepackage{cleveref}

\makeatletter
\providecommand{\@fourthoffour}[4]{#4}
\newcommand\fixstatement[2][\proofname\space of]{%
  \ifcsname thmt@original@#2\endcsname
    \AtEndEnvironment{#2}{%
      \xdef\pat@label{\expandafter\expandafter\expandafter
        \@fourthoffour\csname thmt@original@#2\endcsname\space\@currentlabel}%
      \xdef\pat@proofof{\@nameuse{pat@proofof@#2}}%
    }%
  \else
    \AtEndEnvironment{#2}{%
      \xdef\pat@label{\expandafter\expandafter\expandafter
        \@fourthoffour\csname #1\endcsname\space\@currentlabel}%
      \xdef\pat@proofof{\@nameuse{pat@proofof@#2}}%
    }%
  \fi
  \@namedef{pat@proofof@#2}{#1}%
}

\globtoksblk\prooftoks{1000}
\newcounter{proofcount}

\NewEnviron{proofatend}{%
  \edef\next{%
    \noexpand\begin{proof}[\pat@proofof\space\pat@label]%
    \unexpanded\expandafter{\BODY}}%
  \global\toks\numexpr\prooftoks+\value{proofcount}\relax=\expandafter{\next\end{proof}}
  \stepcounter{proofcount}}

\def\printproofs{%
  \count@=\z@
  \loop
    \the\toks\numexpr\prooftoks+\count@\relax
    \ifnum\count@<\value{proofcount}%
    \advance\count@\@ne
  \repeat}
\makeatother

\fixstatement{claim}

\newtheorem{definition}{Definiton}
\newtheorem{example}{Example}
\newtheorem{remark}{Remark}
\newtheorem{theorem}{Theorem}
\newtheorem*{theorem*}{Theorem}
\newtheorem*{claim*}{Claim}

\newcommand{\footremember}[2]{%
    \footnote{#2}
    \newcounter{#1}
    \setcounter{#1}{\value{footnote}}%
}

\author{%
  Ze Chen\footremember{c}{Maximus Labs, ze@maximuslabs.org}%
  \\\and Ruichao Jiang\footremember{j}{Maximus Labs \& Carleton University, jiang@maximuslabs.org}%
  \\\and Javad Tavakoli\footremember{ja}{University of British Columbia, javad.tavakoli@ubc.ca}%
  \\\and Yiqiang Zhao\footremember{z}{Carleton University, zhao@math.carleton.ca}
  }

\bibliography{biblio.bib}

\title{Robbed withdrawal}
\date{\today}
\begin{document}
\maketitle
\begin{abstract}
    In this article we show that Theorem 2 in \citet{wombat} is incorrect. Since Wombat Exchange, a decentralized exchange, is built upon \citet{wombat} and Theorem 2 is fundamental to Wombat Finance, we show that an undesirable phenomenon, which we call the \emph{robbed withdrawal}, can happen as a consequence.
\end{abstract}
\section{Introduction}
Decentralized exchange plays an important role in decentralized finance, where the market-making is not done by an order book but by Automated Market Makers (AMMs). A pool in an AMM is a pair of two tokens. AMMs like Uniswap v2 is double-sided, where the liquidity provider must provide both tokens of the pool. Wombat \citep{wombat} uses Single-Sided Automated Market Maker (SSAMM), where a liquidity provider is allowed to provide only one kind of the token in the pool. Theorem 2 of \citep{wombat} is the backbone of Wombat, which stipulates the equilibrium states of pools in Wombat. However, we refute Theorem 2 of \citet{wombat}, i.e. we not only point out the gap of the proof of Theorem 2 but also 
\begin{enumerate}
    \item directly prove that Theorem 2 is false,
    \item provide a concrete counterexample.
\end{enumerate}
The organization of this article is as follows. We first introduce Wombat's SSAMM in \Cref{sec:wombat-ssamm} to fix the notation. We show that the proof of Theorem 2 in \citet{wombat} is flawed and that the statement of it is false in \Cref{sec:wombat-mistake}.
\section{Wombat SSAMM}
\label{sec:wombat-ssamm}
We fix the notation. This section does not aim to be a comprehensive introduction to AMM nor SSAMM. For those, see \citet{wombat, xu}.
\begin{definition}[Wombat SSAMM]
\label{def:wombat}
    The SSAMM of Wombat is defined by the following equation.
    \begin{equation*}
        F(A_1,L_1,A_2,L_2)=\frac{A_1^2-cL_1^2}{A_1}+\frac{A_2^2-cL_2^2}{A_2}+(c-1)(L_1+L_2)=0,
    \end{equation*}
    where $A_i\geq0$ ($L_i\geq0$), $i=1,2$, is the asset (liability) for token $i$ in the liquidity pool.
\end{definition}
\begin{remark}
    \Cref{def:wombat} uses different variables from that used in \citet{wombat}, where they used $r_i\coloneqq\frac{A_i}{L_i}$ and $L_i$ as the state variables of the SSAMM, which is a reparametrization. \citet{wombat} also uses a constant $D$, which is equal to $(1-c)(L_1+L_2)$ in our notation. This equality can be proven if the global equilibrium, if existent as claimed by Theorem 1 in \citet{wombat}, of Wombat SSAMM is $r^*=1$. Since the statement of Theorem 2 in \citet{wombat} assumes that $r^*=1$ and in the first line of their proof, they derived $D=(1-A)(L_1+L_2)$, where $A$ is our $c$. The reason we use constant $c$ instead of $A$ is to avoid the confussion with the asset varaible $A_i$.
\end{remark}
There are two types of actions defined on the Wombat SSAMM.
\begin{definition}[Swap]
    Swap token $1$ for token $2$: $A_1\gets A_1+\delta$, $A_2\gets A_2+\delta'$, where $\delta>0$ is given and $\delta'$ is the negative solution of the following equation.
    \begin{equation*}
        F(A_1+\delta, L_1, A_2+\delta', L_2)=F(A_1,L_1,A_2,L_2).
    \end{equation*}
\end{definition}
Swapping token $2$ for token $1$ is defined similarly.
\begin{definition}[Liquidity provision/withdrawal]
    Provide liquidity for token $1$: $L_1\gets L_1+\delta_L$, $A_1\gets A_1+\delta_A$, where $\delta_L$ is given and $\delta_A$ is the solution of
    \begin{equation*}
        F(A_1+\delta_A,L_1+\delta_L,A_2,L_2)=F(A_1,L_1,A_2,L_2).
    \end{equation*}
\end{definition}
\begin{remark}
    It is called the liquidity provision if $\delta_L>0$ and liquidity withdrawal if $\delta_L<0$.
\end{remark}
\section{Disproof and counterexample}
\label{sec:wombat-mistake}
\begin{theorem*}[Theorem 2 in \citet{wombat}]
    Assume that $r^*=1$. If $\delta_L<0$, then $\delta_L\leq\delta_A<0$; if $\delta_L>0$, then $0<\delta_A\leq\delta_L$. Furthermore, in both cases, $\delta_L=\delta_A$ iff $r_i=1$.
\end{theorem*}
The Eqn (13) in the proof of Theorem $2$ in \citet{wombat} reads\footnote{Their constant $A$ is replaced by our $c$ and we omit all subscripts $i$ in their equation.}
\begin{equation}
    \label{eq:wombat-mistake}
    \delta_A-\frac{c(L+\delta_L)^2}{A+\delta_A}+\frac{cL^2}{A}=(1-c)\delta_L.
\end{equation}
Their proof goes as follows.
\begin{quote}
    \emph{``If $\delta_L<0$ and $\delta_A\geq0$, then the left hand side (LHS) of Eq. (13)\footnote{Our Eqn \eqref{eq:wombat-mistake}.} is positive while the right hand side (RHS) is negative, a contradiction."}
\end{quote}
It is obvious that, if $c>1$ and $\delta_L<0$, then the RHS of Eqn \eqref{eq:wombat-mistake} is positive. Their proof is therefore gapped. In fact, $c$ is known as the amplification factor in \citet{wombat} and is only assumed to be greater than zero but not necessarily less than one. As shown in Fig 1 in \citet{wombat}, the amplification factor $c$ is shown to be equal to $300$ for Wombat. Hence our point is not vacuous.

However, this gap does not falsify the statement of Theorem $2$. Perhaps the claim is correct and the proof can be fixed. The following theorem shows that it is not the case.
\begin{theorem}[Robbed withdrawal]
\label{theorem:robbed-withdrawal}
    Let $-L<\delta_L<0$. Under the following three conditions:
    \begin{enumerate}
        \item $3L+\delta_L<A$,
        \item $\delta_L\in\left[-L,-\frac{L^2}{A}\right)$,
        \item $c\in\left(\frac{A}{A-2L-\delta_L},\frac{A}{L}\right)$,
    \end{enumerate}
    One either has
    \begin{itemize}
        \item $\delta_A>0$, or
        \item $\delta_A<0$ and $|\delta_A|>A$\footnote{This implies $|\delta_A|>A>L>|\delta_L|\implies\delta_A<\delta_L$, contradicting the claim $\delta_L\leq\delta_A$ in Theorem $2$ of \citet{wombat}}.
    \end{itemize}
\end{theorem}
\begin{proof}
    The solution to Eqn \eqref{eq:wombat-mistake} is
    \begin{equation*}
        \delta^{\pm}_A=\frac{-b\pm\sqrt{b^2-4\left[(A-2L-\delta_L)c-A\right]\delta_L}}{2},
    \end{equation*}
    where $b=\left(\delta_L+\frac{L^2}{A}\right)c+A-\delta_L$.
    \begin{claim*}
        $2L<b<2A$.
    \end{claim*}
    Assuming that the above claim is true, we have
    $$\delta_A^+>0\iff-\left[(A-2L-\delta_L)c-A\right]\delta_L>0\iff c>\frac{A}{A-2L-\delta_L},$$
    where we used $\delta_L<0$ and Condition $1$: $A-2L-\delta_L>L>0$.  
    
    On the other hand, $b>0$ implies $\delta^-_A<0$. Also,
    \begin{equation*}
        \begin{split}
            |\delta_A^-|>A&\iff\sqrt{b^2-4\left[(A-2L-\delta_L)c-A\right]\delta_L}>2A-b\\
            &\iff b^2-4\left[(A-2L-\delta_L)c-A\right]\delta_L>(2A-b)^2\\
            &\iff(L+\delta_L)^2>0,
        \end{split}
    \end{equation*}
    where we used the Claim $b<2A$ to keep the direction of the inequality correct when squaring.
\end{proof}
Now we prove the claim.
\begin{proof}
    By  Condition $2$: $\delta_L<-\frac{L^2}{A}$ and Condition $3$: $c<\frac{A}{L}$, we have
    \begin{equation*}
        \begin{split}
            b&>\left(\delta_L+\frac{L^2}{A}\right)\frac{A}{L}+A-\delta_L\\
            &=A+L+\frac{A-L}{L}\delta_L>2L>0,
        \end{split}
    \end{equation*}
    where we used $-L<\delta_L$ for the second inequality.
    
    Similarly, by Condition $2$: $\delta_L<-\frac{L^2}{A}$ and Condition $3$: $\frac{A}{A-2L-\delta_L}<c$, we have
    \begin{equation*}
        \begin{split}
            b&<\left(\delta_L+\frac{L^2}{A}\right)\frac{A}{A-2L-\delta_L}+A-\delta_L\\
            &=\frac{A\delta_L+L^2}{A-2L-\delta_L}-\delta_L+A\\
            &=\frac{(L+\delta_L)^2}{A-2L-\delta_L}+A\\
            &<\frac{(L-L^2/A)^2}{A-2L+L^2/A}+A\\
            &=\frac{(1-L/A)^2L^2}{(1-L/A)^2A}+A\\
            &=\frac{L^2}{A}+A<2A,
        \end{split}
    \end{equation*}
    where we used the fact that function $f(\delta_L)=\frac{(L+\delta_L)^2}{A-2L-\delta_L}$ is increasing in $\left[-L,-\frac{L^2}{A}\right)$ in the fourth line. To see it,
    \begin{equation*}
        f'(\delta_L)=\frac{(L+\delta_L)(2A-3L-\delta_L)}{(A-2L-\delta_L)^2}>0\iff-L<\delta_L<2A-3L.
    \end{equation*}
\end{proof}
The following example demonstrates that the robbed withdrawal can happen during a trading process.
\begin{example}
    Let $c=2.008$.
    \begin{itemize}
        \item Initialize: $A_1=L_1=100$, $A_2=L_2=100$.
        \item Swap $101$ token $1$ for token $2$: $A_1\gets201$, $A_2\gets55.98$.
        \item Withdraw $99.1$ token $1$. At this moment, \begin{enumerate}
            \item $3L_1+\delta_L=200.9<201$,
            \item $\delta_L=-99.1\in[-100,-49.75]\subseteq\left[-L_1,-\frac{L_1^2}{A_1}\right)$,
            \item $c=2.008\in(2.00799,2.01)\subseteq\left(\frac{A_1}{A_1-2L_1-\delta_L},\frac{A_1}{L_1}\right)$,
        \end{enumerate}
        all conditions of \Cref{theorem:robbed-withdrawal} are satisfied. Indeed, Eqn \eqref{eq:wombat-mistake} has roots $\delta_A\approx-201.009$ or $0.001$. Then either $A_1\gets-0.009$ (the platform being robbed) or $201.001$ (the liquidity provider being robbed).
    \end{itemize}
\end{example}
\section{Conclusion}
In Wombat, withdrawal of liquidity in token $i$ is always associated with a fee\footnote{It is inappropriate to call this a fee. An infinitesimal amount of it is calculated by an implicit differentiation $\frac{\partial A_i}{\partial L_i}=-\frac{\partial F/\partial L_i}{\partial F/\partial A_i}$, as is the slippage during a swap $\frac{\partial A_j}{\partial A_i}=-\frac{\partial F/\partial A_i}{\partial F/\partial A_j}$. We find liquidity slippage a better name.} unless $A_i=L_i$ \citep{wombat}. \Cref{theorem:robbed-withdrawal} says that under its conditions, one of the following two cases happens
\begin{enumerate}
    \item $\delta_A>0$: After the liquidity provider burns their liquidity tokens, not only do they receive no asset back but also they have to give the platform a positive amount of asset, in other words, the liquidity provider is robbed;
    \item $\delta_A<0$ and $|\delta_A|>A$: The platform must provide more asset to the liquidity provider than it has, in other words, the platform is robbed (also known as a bad debt).
\end{enumerate}
Therefore, we name the above phenomenon the \emph{robbed withdrawal}.
\printbibliography
\end{document}